\documentclass{article}

\usepackage{amssymb,amsfonts,amsmath}
\usepackage{graphicx} 
\usepackage{cite}

\newcommand{\df}{\Delta F}
\newcommand{\pavg}{\overline{\ptarg}}
\newcommand{\pobs}{p^\text{obs}}
\newcommand{\Pobs}{P^\text{obs}}
\newcommand{\ptarg}{p}
\newcommand{\Ptarg}{P}
\newcommand{\tcorr}{t_\text{corr}}
\newcommand{\tsim}{t_\text{sim}}
\newcommand{\wbb}{w^\text{bb}}

\begin{document}

\title{A ``black-box'' re-weighting analysis can correct flawed
simulation data, after the fact}
\author{F.\ Marty Ytreberg\footnote{E-mail:
    ytreberg@uidaho.edu}\\
    Department of Physics,\\
    University of Idaho, Moscow, ID 83844-0903,\\
    Daniel M.\ Zuckerman\footnote{E-mail:
    dmz@ccbb.pitt.edu}\\ Department of Computational Biology,\\
    University of Pittsburgh, 3501 Fifth Avenue, Pittsburgh, PA 15260}
\maketitle

\begin{abstract}
There is a great need for improved statistical
sampling in a range of physical, chemical and biological systems.
Even simulations based on correct algorithms suffer from statistical error, 
which can be substantial or even dominant when slow processes are involved.
Further, in key biomolecular applications, such as the determination of 
protein structures from NMR data, non-Boltzmann-distributed
ensembles are generated.
We therefore have developed the ``black-box'' strategy for re-weighting a 
set of configurations \emph{generated by arbitrary means} to produce an 
ensemble distributed according to any target distribution.
In contrast to previous algorithmic efforts, the black-box
approach exploits the configuration-space density \emph{observed} in a 
simulation, rather than assuming a desired distribution has been generated.
Successful implementations of the strategy, which reduce
both statistical error and bias, are developed for a one-dimensional
system, and a 50-atom peptide, for which the correct 250-to-1
population ratio is recovered from a heavily biased ensemble.
\end{abstract}

Ensemble averages over configurations are fundamental to
the analysis of finite-temperature 
systems of physical, chemical, and biological interest, as well as to any
statistically defined system.
Yet it is well appreciated that estimates of such averages based
on computer simulations can suffer from both systematic
and statistical error \cite{landau-book,frenkel-book}.
We therefore ask:
\emph{Given a set of previously generated configurations of uncertain quality,
what is the best way to estimate ensemble averages?}
Our proposed answer, the ``black-box re-weighting'' (BBRW) strategy
described below, appears promising in its ability to overcome
both types of error in some systems.

Statistical error is a ubiquitous problem of
under-appreciated practical importance.
Every algorithm known to the authors, including sophisticated methods
\cite{swendsen-repx,berg-muca,lyman-resx},
relies on repeated visits to
a state (a subset of configuration space) in order to generate
statistical reliability or precision in the population estimate for that state.
If we define the simulation-and-system-specific correlation time
$\tcorr$ as the time required to visit all important states at
least once, then statistical precision requires a long total
simulation time, $\tsim \gg \tcorr$.
Standard square-root-of-duration arguments \cite{landau-book,frenkel-book}
suggest that a simulation retains a fractional
imprecision of $\sqrt{\tcorr/\tsim}$ (on a unit scale).
Below, we show that BBRW dramatically
cuts statistical error, avoiding the slow square-root behavior.

Systematic bias is typical in some systems of great practical
importance---such as full-sized proteins---where, to date, 
it is not clear that any
atomically detailed simulation has come close to reaching $\tcorr$.
Indeed, surprisingly lengthy simulations are required to obtain
statistical ensembles for small peptides \cite{lyman-converge}.
Nevertheless, biased non-Boltzmann distributed
sets of atomically detailed protein structures
\emph{are} regularly generated, e.g., for
NMR structure determination \cite{spronk} and in the study of protein
folding \cite{micheletti}.
These sets can be useful for docking \cite{knegtel},
which is employed in drug design.
The proposed BBRW strategy, in principle,
can convert such sets into statistically distributed ensembles.

\section{Background and theory}

The fundamental basis of our re-weighting
approach is the recognition that a
simulation's output (the configurations generated) \emph{contains
valuable information that is not conventionally used} to estimate
ensemble averages. This information is the
observed configuration-space density.
It is critical to note that the observed density never matches the
targeted distribution (e.g., proportional to a Boltzmann factor)
because of
(i) the omnipresent statistical error, and also possibly,
(ii) errors in the algorithm or software.
Our approach can treat both statistical and systematic error because
it is blind to the source of error. 

\subsection{Theory}\
Standard re-weighting theory assumes that configurations in
a simulated ensemble are already distributed according to a known distribution, typically proportional to a Boltzmann factor
\cite{swendsen-ferrenberg,liu-book,okabe,clementi-reweight,escobedo}.
The novelty of our re-weighting method
is that \emph{we treat configurations as if they were generated by
an unknown process}, a ``black box.''
Thus, any ensemble can be treated since the results of the
re-weighting do not depend on the specific process that was used to
generate the ensemble.

Our goal is to estimate ensemble averages and
relative weights for configurations $\vec{x}_j$---denoted by $j$---in
the target distribution $\Ptarg(j)$,
regardless of how the configurations were generated.
While our approach will be general, we will assume for concreteness that 
the target ensemble is a classical system
with potential energy $U(\vec{x})$.
In this case, the targeted probability density is 
    $\Ptarg(j)=Z^{-1}e^{-U(\vec{x}_j)/k_BT}$,
where $Z = \int d\vec{x} e^{-U(\vec{x})/k_BT}$
is the configurational partition function at temperature $T$.
Because $Z$ is typically not known, it will prove useful
to define the un-normalized distribution
\begin{equation}
\ptarg(j) = e^{-U(\vec{x_j})/k_BT} \propto \Ptarg(j) \, .
\label{ptarg}
\end{equation}
The overall partition function $Z$ will not affect our
quest to assign \emph{relative} configurational weights.

The sample to be analyzed consists of a set of $N$ configurations 
$\{ \vec{x}_1, \vec{x}_2, \ldots, \vec{x}_N \}$
which may have been generated by an arbitrary simulation protocol.
This set of configurations is distributed according to the
\emph{observed} probability density $\Pobs(j)$, by definition. 
In general the observed and targeted distributions will differ,
$\Pobs(j) \neq \Ptarg(j)$, 
due to statistical and/or systematic error as explained earlier.
It will again prove useful to consider an un-normalized density denoted by
$\pobs(j) \propto \Pobs(j)$.

With the targeted and observed distributions defined,
the straightforward logic of standard re-weighting \cite{liu-book}
may be applied.
In an ideal sample, an infinitesimal configuration-space volume $d\vec{x}$
about $j$ would contain a quantity of configurations proportional to
the targeted distribution $\ptarg(j) \, d\vec{x}$.
However, in actuality, $\pobs(j) \, d\vec{x}$ is obtained.
The erroneous sampling can be corrected to cancel the error, 
by applying a \emph{relative} weight
\begin{equation}
    \wbb(j)=\ptarg(j)\big/\pobs(j)
    \label{eq-bbw}
\end{equation}
to each configuration.
For the physical systems considered here
the target distribution is given exactly by the
standard Boltzmann factor of Eq.\ \eqref{ptarg}.
We therefore assume $\ptarg$ is ``known'' in the sense that it
can be evaluated for an arbitrary configuration. 
Then, because Eq.\ \eqref{eq-bbw} is exact, the key to practical
application of this equation is accurate estimation of $\pobs$.

Note that the relative weights of Eq.\ \eqref{eq-bbw}
will only be applied to the $N$ configurations already
generated---and therefore do not
report on regions of configuration space not sampled. 

Ensemble averages of any function $f$, in our approach, are
given by a weighted sum over sampled configurations $\vec{x}_j$:
\begin{equation}
    \langle f \rangle_\text{bb} =
	\displaystyle\sum_{j=1}^N \wbb(j) \, f(j) \bigg/
	\displaystyle\sum_{j=1}^N \wbb(j) \, .
    \label{eq-bbave}
\end{equation}
The average \eqref{eq-bbave}
differs from what we term a ``canonical average''
of the same configurations, where one assumes 
$\pobs(j) \propto \ptarg(j)$;
then $\wbb(j)$ is a constant, and averages 
reduce to the usual un-weighted estimate
$\langle f \rangle_\text{can} = \sum_{j=1}^N f(j)/N$ \cite{frenkel-book}.
Note that the proportionality constants
for $\ptarg(j)$ and $\pobs(j)$ cancel in \eqref{eq-bbave}
and thus do not affect the average. 

To see why it is better to re-weight samples when $\pobs$
can be estimated well, consider the extreme case of a
discrete system with six equi-probable states $s$ (e.g., a cubic die).
Of course, all ensemble properties can be calculated by hand,
but what if statistical sampling were relied upon?
Assume $N=30$ samples were generated (either by a biased or
unbiased procedure), yielding the frequencies $(8,4,2,4,7,5)$,
proportional to $\pobs(s)$ for $s = (1, \ldots, 6)$, respectively.
Clearly, the canonical assumption $\pobs \propto \ptarg$
will lead to substantial errors in ensemble averages.
However, the frequencies of observation will
\emph{exactly cancel} $\pobs$ when the relative weights $\wbb$ are used in
\eqref{eq-bbave}---summing over all 30 ``configurations''---to yield
exact ensemble averages.
The extension of this simple idea to continuum systems
is the subject of this report. 

Several additional points should be made regarding the
strategy embodied in \eqref{eq-bbw} and \eqref{eq-bbave}:
(i) It is an \emph{analysis} scheme, treating the same set of
configurations as might be analyzed canonically;
the approach is not capable of predicting
populations for regions of configuration space that are not
in the observed ensemble.
(ii) The observed relative probabilities $\{\pobs(j)\}$ will \emph{always}
differ from those intended (e.g., canonical Boltzmann factors)
due to statistical error---and perhaps significantly.
(iii) The relative weights \eqref{eq-bbw} are valid 
regardless of the specific process used to generate configurations.
(iv) The approach is particularly suited to estimating
conformational free energy differences, as described below.
(v) As in the single histogram approach \cite{swendsen-ferrenberg},
the BBRW method retains the ability to re-weight configurations
into an arbitrary target ensemble $\ptarg$.
(vi) The principal practical challenge in implementing our
strategy is the estimation of $\pobs$.

\subsection{Free energy differences}\
To provide a proof of principle for our re-weighting
approach we will determine free energy differences ($\df$),
between states for various test systems
\cite{karplus-deca,honig-colony,gilson-jacs,meirovitch-argon,meirovitch-deca,ytreberg-absf,ytreberg-shift}.

A well established canonical technique for determining $\df$ is to
use ordinary simulation methods (e.g., molecular dynamics)
to generate an ensemble which is distributed according to
the desired distribution $\ptarg$.
Then, the free energy difference is defined as
\begin{equation}
    e^{-\df/k_BT} \equiv Z_B \big/ Z_A = N_B \big/ N_A \, ,
    \label{eq-can-df}
\end{equation}
where $Z_s = \int_{\vec{x} \, \in \, s}d\vec{x} \, e^{-U(\vec{x})/k_BT}$
is the configurational partition function for state
$s$ with coordinates given by $\vec{x}$,
and $N_s$ is the number of configurations observed in state $s$.

For our re-weighting approach, we do not assume that
configurations are distributed according to $\ptarg$.
Instead, each configuration must first be assigned a relative weight
using Eq.\ \eqref{eq-bbw};
then $\df$ is estimated based on summing
these weights in the respective states:
\begin{equation}
    e^{-\df/k_BT} \equiv Z_B\big/Z_A =
	\displaystyle\sum_{j \, \in \, B} \wbb(j) \bigg/
	\displaystyle\sum_{j \, \in \, A} \wbb(j) \, ,
    \label{eq-bb-df}
\end{equation}
where $\sum_{j \, \in \, s}$ represents a sum over all relative
weights $\wbb(j)$
in the state $s$.

\subsection{One-dimensional thought experiment}\
The central idea behind our re-weighting approach is that one can
utilize the \emph{observed} configuration-space density
to re-weight an arbitrary set of configurations into any desired ensemble.
Because this idea is so fundamental to the present discussion,
and apparently novel, 
it is useful to illustrate the idea using a one-dimensional
potential.

\begin{figure}
    \begin{center}
	\includegraphics[width=5cm,clip]{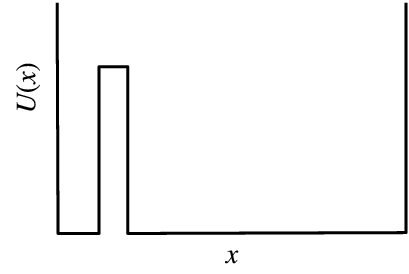}
    \end{center}
    \caption{
	One-dimensional potential
	for a ``thought experiment.''
        The energy is constant and equal in both states, but the right
        state is 10 times as wide as the left.
	\label{fig-square}
    }
\end{figure}

Consider the equal-energy double-square-well system depicted in
Fig.\ \ref{fig-square}, for which we would like to estimate the
relative populations of the two states.
Assume that the barrier is so high that it cannot be crossed
in a conventional simulation of affordable length, but
that we can run independent simulations in each state
(a common situation for biomolecular systems when
multiple structures are known). 
The states' widths are in the ratio 1:10. 

After generating, say, equi-sized trajectories in each state, a
conventional view would hold that no analysis of the relative
populations is possible since the barrier has not been crossed.
However, since the right well is ten times wider than the left,
the observed configuration-space density in the right
well will be, on average, one-tenth that in the left.
Combined with knowledge of the two ensemble sizes, this is sufficient
information for calculating the state populations.

More explicitly, one can estimate the observed density by counting
configurations in equi-sized histogram bins, leading to
$\pobs(j) \propto n_b(j)$, where $n_b(j)$ is the count in the
bin containing configuration $j$.
In this example, $\ptarg(j) = \text{const}$ for all $j$ because the
energy is constant over both states.
Now, by using Eq.\ \eqref{eq-bbw} in Eq.\ \eqref{eq-bb-df},
the sum over relative weights leads to \emph{exact} cancellation of
$n_b$ for each bin.
Thus, this purely entropic case is correctly handled,
ultimately yielding simply the ratio of the numbers of (equi-sized)
bins in each state---i.e., the ratio of state widths.
Below, we will see that the ``easier'' energetic effects are
also treated correctly in less trivial examples.

The key point is that \emph{the observed density $\pobs$, 
combined with knowledge of the desired ensemble $\ptarg$,
provides sufficient information to deduce the state populations,
regardless of the sampling bias}.  
Perhaps equally importantly, we note that the density can be estimated
in different ways besides binning, e.g., using configuration-space
distances between nearby configurations. 

The reasoning used to correct systematic bias can be applied
equally to statistical error.
Assume now that we possess a single
continuous trajectory which has crossed the barrier in Fig.\ \ref{fig-square} 
at least once.
The population estimates will be statistically flawed due to finite sampling. 
Yet, our re-weighting analysis will still yield
correct results since the approach does not depend
on the source of the bias in the data.

Below we will show that such density information is both present and
often accessible in high-dimensional systems.  We emphasize that the
information is always present in principle and does not depend on prior
knowledge.

\section{Results and discussion}

How can the observed density be computed in a way that will be useful,
even for high-dimensional systems?
We will explore two distinct strategies: binning
and also the use of configuration-space distances
(``nearest-neighbors strategy'').
For high-dimensional systems, we will demonstrate that
not all coordinates need be considered in the analysis.
We emphasize that other strategies for estimating density are
possible \cite{silverman-book,scott-book}.

The test systems chosen for this study are a one-dimensional double-well
potential and a 50-atom dileucine peptide
(ACE-(leu)$_2$-NME).
Our reasoning for choosing these systems are:
(i) The system sizes are small enough that it is possible to
compute an independent $\df$ estimate via counting,
i.e., using Eq.\ \eqref{eq-can-df}.
(ii) Both are non-trivial: the one-dimensional potential has
high barrier, and dileucine has large side chains and thus
144 degrees of freedom.

\subsection{Binning strategy}\
In the simplest use of bins, the observed density is estimated based on
simple counting: omitting normalization, $\pobs(j) = n_b(j)$,
where $n_b(j)$ is the number of configurations in
the bin which includes configuration $j$.
Indeed we have tried this scheme and it works, but not optimally.

A more effective procedure for weighting individual configurations
combines simple counting with the assumption
of local equilibration: within bins, configurations are assumed
distributed according to $\ptarg$ itself.
This assumption could correspond to the case where separate canonical
simulations have been performed in different regions of
configuration space, or to a canonical simulation which was not
equilibrated on long time scales but well-sampled locally.
The local-equilibration estimate for the observed distribution is
\begin{equation}
    \pobs(j) = n_b(j) \, \ptarg(j) \big/ \, \pavg_j \, , 
    \label{eq-Pbin}
\end{equation}
which implies $\wbb(j) = \pavg_j\big/n_b(j)$.
The bin-specific normalization is the average relative target
probability in the bin,
namely, $\pavg_j = \sum_{i \in \text{bin}} \ptarg(i) / n_b(j)$.
This normalization serves to keep the total observed probability of
a bin proportional to the number of counts,
and independent of the local Boltzmann factors. 
As explained in the Supplementary Material,
a consequence of adopting the approximation \eqref{eq-Pbin},
is that $\pavg_j$ is implicitly assumed 
proportional to the \emph{local} partition function for the bin
containing configuration $j$.
(Each bin has a different local partition function estimate.)
Nevertheless, we emphasize that the use of local partition
functions is not intrinsic to the black-box approach,
but only to binning methods for estimating $\pobs$:
see, for instance, the nearest-neighbor strategy described below.

\begin{figure}
    \begin{center}
	\includegraphics[height=3cm,clip]{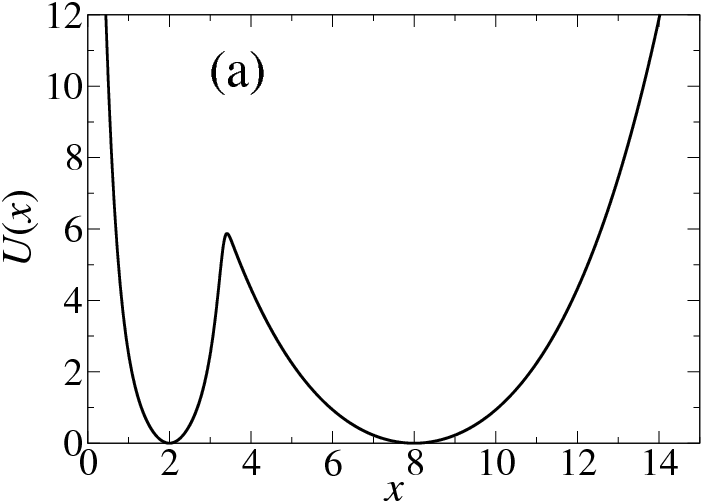}
	\includegraphics[height=3cm,clip]{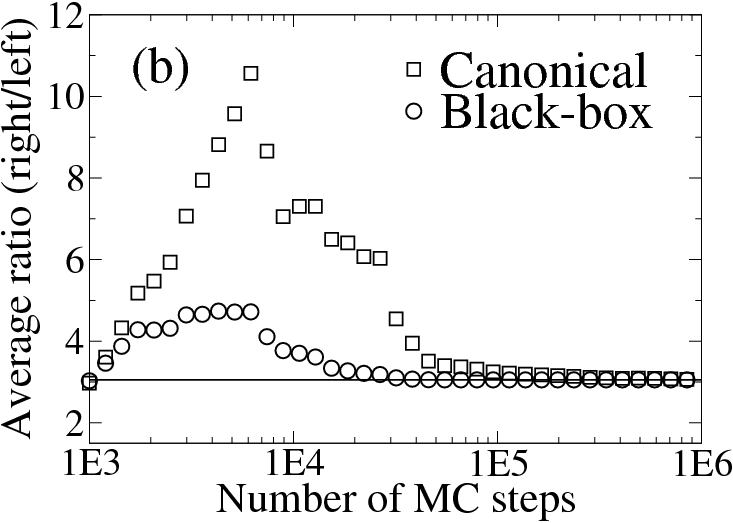}
	\includegraphics[height=3cm,clip]{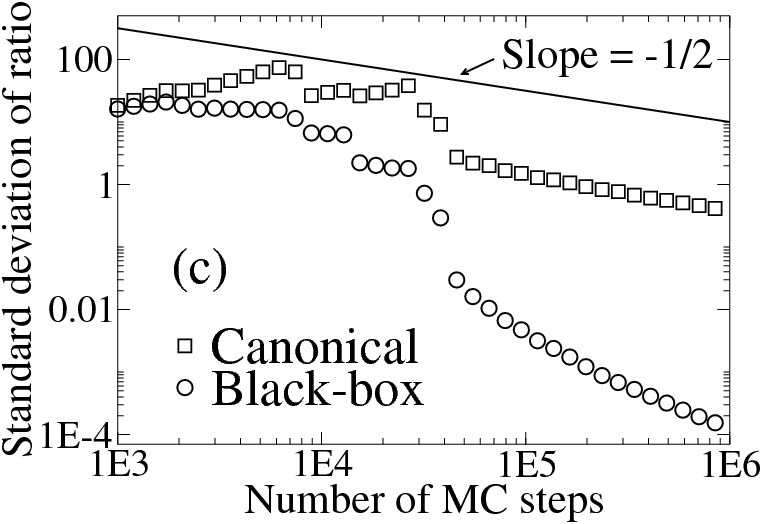}
	\includegraphics[height=3cm,clip]{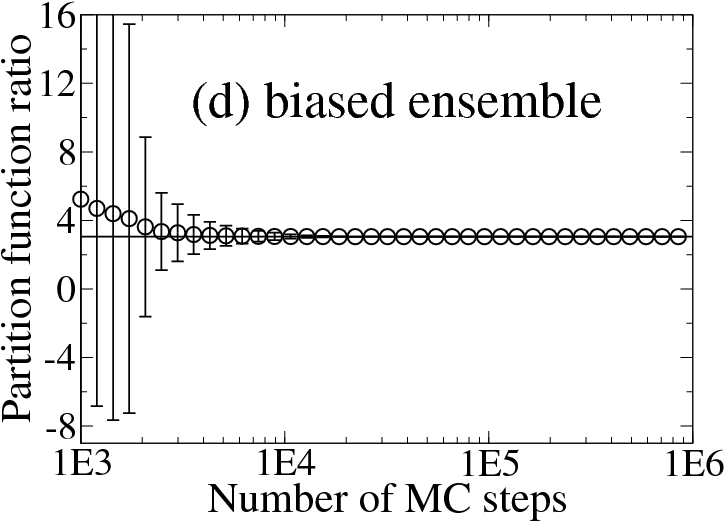}
    \end{center}
    \caption{
	Results for a one-dimensional double-well test system.
	(a) The potential energy plotted as a function of the coordinate.
	(b) The ratio of partition functions between the left and right wells.
	All results shown are the average
	of $10^3$ independent computations.
	The canonical estimates (squares) are given by
	Eq.\ \eqref{eq-can-df}, the black-box estimates (circles)
	by Eq.\ \eqref{eq-bb-df}.
	An independent estimate, obtained via integration,
	is shown by the solid horizontal line.
        (c) The standard deviation from the same data as
	that for figure (b) is plotted showing the rapid
	decrease in the standard deviation for our method
	(circles) as compared to canonical (squares).
	(d) Estimate of the ratio of partition functions
	using our re-weighting approach via Eq.\ \eqref{eq-bb-df},
	based on a heavily biased ensemble of configurations.
	Note that canonical estimation using Eq.\ \eqref{eq-can-df}
	is not possible.
	\label{fig-1d}
    }
\end{figure}

\emph{One-dimensional double-well potential.}
Here we consider the smooth potential of Fig.\ \ref{fig-1d}a.
The relative target probability is defined by  
$\ptarg(x) = e^{-U(x)/k_BT}$, where $U(x)$ is the potential
energy sketched for the approximately 6 $k_BT$ barrier height used here. 
The barrier heights for this potential, as
well as basin curvatures and separations, are fully adjustable, using,
$U(x) = -\ln\big[\exp\big(-0.5(x-x_a)^6 -(x-x_a)^2/2s_a^2\big)
    +\exp\big(-0.003(x-x_b)^4-(x-x_b)^2/2s_b^2\big)\big]$,
where we have used $x_a=2.0$, $x_b=8.0$, $s_a=0.5$, and $s_b=1.5$.
We estimated the ratio of partition functions
for the two wells (right over left) using data obtained from ordinary
canonical simulation, i.e., Metropolis Monte Carlo (MC) \cite{metropolis}.

Fig.\ \ref{fig-1d} shows the differences between
re-weighting, using a binning strategy,
and canonical estimates of the ratio
of partition functions between two wells (right over left) in (a).
Results shown are the averages (b)
and standard deviations (c) from $10^3$ independent computations.
The canonical estimates (squares) were computed via \eqref{eq-can-df},
and the black-box estimates (circles) via Eqs.\ \eqref{eq-bb-df} and
\eqref{eq-Pbin} with a bin size of 0.005.
\emph{All data were generated using the same MC trajectories,}
each initiated from the left basin of the potential in (a).
Specifically, all MC trajectories were generated by starting the
system at $x=2.0$.  Each trajectory consisted of $10^6$ trial moves
generated by adding a uniform random deviate between -1.0 and 1.0 to the
current position.
An independent estimate, obtained via numerical integration,
is shown by the solid horizontal line in (b).
Note that in (b) the apparent accuracy of canonical estimation
at short times represents only the average behavior;
the statistical uncertainty is so large that any individual
estimate would be completely unreliable.

Figure \ref{fig-1d}c demonstrates that the error in our re-weighting
analysis decreases at a rate greater than the typical square-root
behavior, as seen for the canonical error.
Since the same ensembles were used for both the re-weighting and canonical
estimates, this improvement may lead one to believe we are getting
something for nothing.
In fact, there are costs, but in essence, they have already been paid:
(i) The re-weighting approach requires knowledge of the desired
distribution; in our case (and most other cases of interest)
this is given by the Boltzmann factor.
(ii) The observed density 
is available information that is not generally used for
analysis of simulation data, and requires only a small
additional computational cost.

In Fig.\ \ref{fig-1d}d we show that our re-weighting approach
can be used to estimate the correct populations between
the left and right wells,
with \emph{an example that simply cannot be treated canonically.}
We generated a biased ensemble,
with configurations that were \emph{not} distributed
according to $\ptarg$ (Boltzmann factor for $U$ shown in (a)).
Then Eqs.\ \eqref{eq-Pbin} and \eqref{eq-bb-df} were
used with a bin size of 0.005 to estimate the ratio
of partition functions.
The data show the averages (circles)
and standard deviations (error bars) for
$10^3$ independent computations.
Since our re-weighting approach is independent of the method
used to generate the ensemble,
we are able to obtain the correct ratio, even
though the ensemble used in the analysis was heavily biased.
For completeness,
we note that the biased ensemble configurations were generated
using the same MC protocol as above,
but with a square well potential defined as
$U(x)=0$ between $x=0$ and $x=15$ and infinite elsewhere.

\emph{Dileucine peptide.}
In this high-dimensional molecular example, we again apply our approach
to a biased ensemble, where canonical estimation would be meaningless.
We use the method to estimate the
free energy difference $\df$ between conformational states
of the 50-atom dileucine peptide
based on independent simulations of each state. 
The two states considered were defined by backbone dihedrals
angles---``alpha'':
$-140.0<\Phi_{1,2}<-80.0 \;\text{and}\; -120.0<\Psi_{1,2}<-60.0$;
and, ``beta'':
$-125.0<\Phi_{1,2}<-65.0 \;\text{and}\; 120.0<\Psi_{1,2}<180.0$.

\begin{figure}
    \begin{center}
	\includegraphics[height=3.5cm,clip]{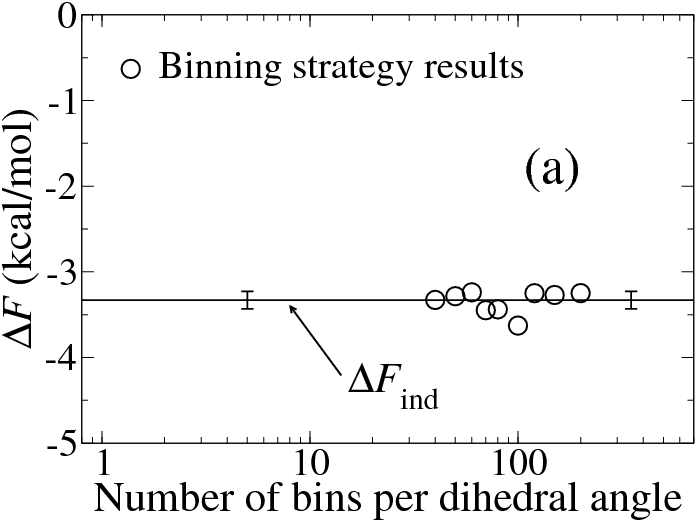}
	\hspace{24pt}
	\includegraphics[height=3.5cm,clip]{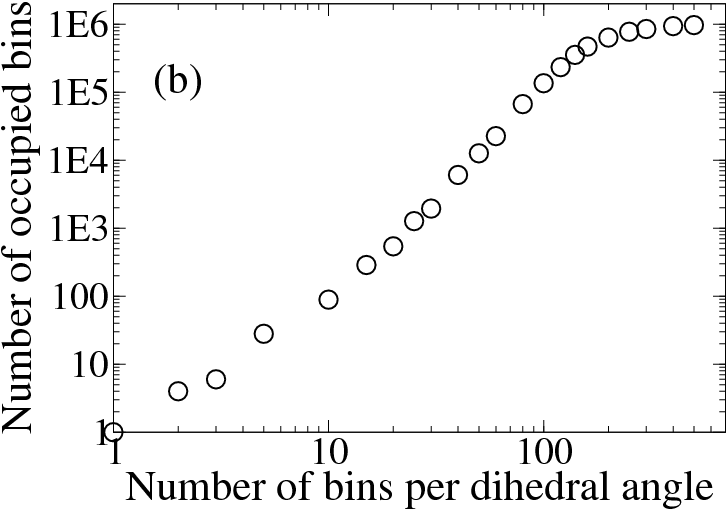}
    \end{center}
    \caption{
	(a) The free energy difference between
	the alpha and beta states of dileucine from binning.
	Results from our re-weighting approach using a heavily
	biased ensemble are shown.
	Data were generated using Eqs.\ \eqref{eq-bbw},
	\eqref{eq-bb-df} and \eqref{eq-Pbin},
	using only the backbone dihedrals $\Phi$ and $\Psi$.
        The independent estimate is shown by the solid horizontal line
	with an approximate error determined by block averaging
	\cite{flyvbjerg}.
	(b) Box-counting analysis of dileucine configurations
        used to determine the range of bin sizes for figure (a).
	\label{fig-bin}
    }
\end{figure}

To test the accuracy of the results from our re-weighting approach,
we also generated an independent estimate of $\df$.
An ordinary, unrestrained, 1.0 $\mu$s simulation was performed and analyzed
via Eq.\ \eqref{eq-can-df}, yielding
$\df_\text{ind} \approx -3.3 \pm 0.1$ kcal/mol, with
statistical error estimated via block averaging \cite{flyvbjerg}.
The simulation data were generated using
{\sc tinker} 4.2 with the OPLS-AA forcefield \cite{oplsaa}.
A 1.0 fs timestep was utilized, and the
simulation was coupled to a 300.0 K bath using
the Andersen thermostat \cite{andersen}.
Solvation was provided by the generalized Born surface area (GBSA)
implicit model \cite{still}.

To generate a BBRW estimate based on the locally equilibrated
scheme of Eq.\ \eqref{eq-Pbin}, simulations of dileucine were performed
constrained to either the alpha or beta state.
We generated 500,000 configurations in each state based on a total
of $\approx$104 ns.
Bins used in \eqref{eq-Pbin} were constructed uniformly using 
only the four backbone
dihedrals---out of 144 degrees of freedom---leading
to a four-dimensional histogram.
Note that empty bins, which are expected for physical and statistical reasons,
simply do not contribute to averages computed via Eq.\ \eqref{eq-bbave},
since only sampled configurations are summed over.

Figure \ref{fig-bin}a shows that our re-weighting approach
can be used to correctly predict the 
free energy difference between the alpha and beta states of dileucine.
The solid horizontal line is the independent estimate
$\df_\text{ind}$ computed via Eq.\ \eqref{eq-can-df}
with an approximate error determined by block averaging \cite{flyvbjerg}.
The circles are results from our re-weighting method using the
backbone dihedral angles $\Phi$ and $\Psi$.
These data were computed using Eqs.\ \eqref{eq-bbw}, \eqref{eq-bb-df},
and \eqref{eq-Pbin}, and plotted as a function of the
number of bins used per dihedral angle in the histograms.
Even though an equal number of alpha
and beta configurations were utilized in our analysis,
the ratio of $\approx 250$ is recovered.
Note that our strategy, as shown here, is an
example of an end-point free energy difference method,
since no path connecting alpha and beta is needed
(e.g., Refs.\ \cite{karplus-deca,honig-colony,gilson-jacs,meirovitch-argon,meirovitch-deca,ytreberg-absf,ytreberg-shift}).

Fig.\ \ref{fig-bin}a also demonstrates a certain
robustness in the approach:
there is a broad range of bin sizes that can be used generate
the correct free energy difference.

Two criteria were used to determine the range of bin sizes
used for the re-weighting results in Fig.\ \ref{fig-bin}a:
(i) The range of bin sizes for accurate $\df$ estimation
apparently 
corresponds to the power-law regime of a plot for estimating the
box counting dimension \cite{strogatz-book}.
This plot is shown in Fig.\ \ref{fig-bin}b
where the linear regions denote the power-law behavior.
(ii) The extreme bin size results are known exactly.
If one bin per dihedral angle is chosen, then the estimate
will always be poor and $\df=0$ (for equi-sized ensembles).
At the other extreme, if a very large number of bins per
dihedral angle is used, then each bin will have a count of 1,
and $\pobs=1$ for every structure.
Since no true density information is obtained near either extreme,
we excluded such estimates.

\subsection{Nearest-neighbors strategy}\
In addition to the binning implementation above,
we now estimate $\pobs$ of Eq.\ \eqref{eq-bbw}
by computing distances between conformations.
The motivation behind pursuing such non-binning approaches
is the realization that for biomolecular systems, it may prove
difficult to populate the necessary high-dimensional histogram bins.
The number of bins will increase exponentially with the number of
coordinates, but the density of sampling in configuration space can
always be estimated based on ``distances'' in configuration space.

In general, to implement the nearest-neighbors strategy \cite{hnizdo-nn},
the distances between all configurations
in the ensemble are computed,
using any metric which preserves the Cartesian volumes
intrinsic to partition functions.
By definition, the dimensionality of the metric can be as large as
the number of degrees of freedom needed to describe a conformation.
The observed configuration-space (relative) density for structure
$j$ is then given by
\begin{equation}
    \pobs(j) = \frac{N_\text{dist}(j)}{R_\text{hyp}(j)^d} \, ,
    \label{eq-Pnobin}
\end{equation}
which implies 
$\wbb = \ptarg(j)\,R_\text{hyp}(j)^d / N_\text{dist}(j)$.
Here, $R_\text{hyp}(j)$ is the radius of the hypersphere
centered on structure $j$ enclosing $N_\text{dist}(j)$ structures,
and $d$ is the dimensionality of the
distance metric used.
Typically, as we will demonstrate below, $d$ can be chosen to be
much smaller than the dimensionality of the system.
See Supplementary Material for further discussion of
the approximations entailed when $d$ is less than the
full dimensionality of the system. 

We used Eq.\ \eqref{eq-Pnobin}
to estimate $\pobs$ for a particular structure $j$ using the following
implementation:
(i) Compute the distance (using any appropriate metric)
between structure $j$ and
all other structures in the ensemble.
(ii) Choose a value of $N_\text{dist}$.
The radius of the hypersphere $R_\text{hyp}$ is given
by the distance to the $N_\text{dist}$ nearest neighbor.
For example, if $N_\text{dist}=10$ is chosen,
then $R_\text{hyp}$ is given by the
distance to the tenth nearest neighbor.

\emph{Dileucine peptide.}
We tested the approach embodied in Eq.\ \eqref{eq-Pnobin}
by estimating the free energy difference ($\df$) between the alpha and
beta conformations of dileucine,
using the same restrained (thus biased) ensemble that was used
above.
Following the previous success of using only the torsion angles
of dileucine, we chose to use a dihedral angle distance
between structures $j$ and $k$ defined by
\begin{equation}
    D_{jk} = \sqrt{\;\sum_{l=1}^d (\phi_j^l-\phi_k^l)^2}\, ,
    \label{eq-dist}
\end{equation}
where $\phi^l$ could be any 
dihedral angle (e.g., $\Phi_1,\;\Psi_2,\;\chi_1$),
and $d$ is the number of angles used in the computation,
i.e., the metric dimension.

\begin{figure}
    \begin{center}
	\includegraphics[height=3.5cm,clip]{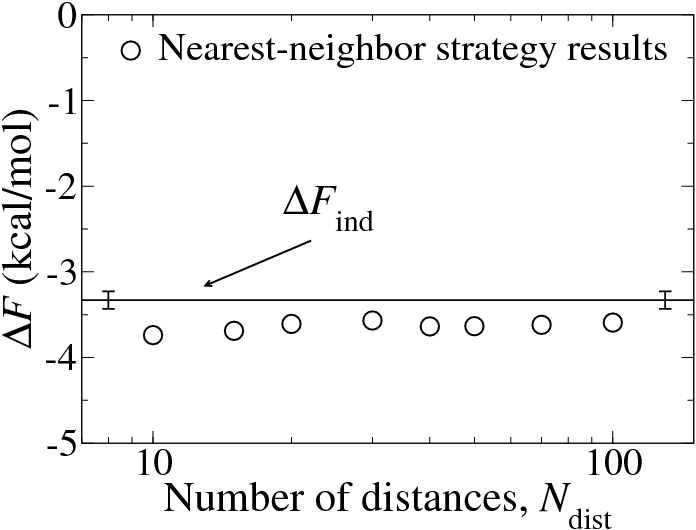}
    \end{center}
    \caption{
	The free energy difference ($\df$) between
	the alpha and beta states of dileucine from a
	nearest-neighbors strategy.
	Results from our re-weighting approach using a heavily
	biased ensembles are shown.
	Data were generated using Eqs.\ \eqref{eq-bbw},
	\eqref{eq-bb-df} and \eqref{eq-Pnobin},
	with distances computed via Eq.\ \eqref{eq-dist}.
	The analysis used only the backbone dihedrals $\Phi$ and $\Psi$.
	Data are given as a function of the number of distances (neighbors)
	used in the analysis.
        The independent estimate is shown by the solid horizontal line
	with an approximate error determined by block averaging
	\cite{flyvbjerg}.
	\label{fig-nobin}
    }
\end{figure}

Figure \ref{fig-nobin} shows that our re-weighting
approach, using the nearest-neighbors strategy, can be used to predict
the free energy difference between the alpha and beta states
of dileucine.
The solid horizontal line is the independent estimate
$\df_\text{ind}$ computed via Eq.\ \eqref{eq-can-df}
with an approximate error determined by block averaging \cite{flyvbjerg}.
The circles are results from our re-weighting method using the
backbone dihedral angles $\Phi$ and $\Psi$.
These data were computed using Eqs.\ \eqref{eq-bbw}, \eqref{eq-bb-df},
and \eqref{eq-Pnobin}, with distances given by Eq.\ \eqref{eq-dist}.
The results are plotted as a function of the number of neighbors
used in the analysis procedure, $N_\text{dist}$.
Even though an equal number of alpha
and beta configurations were utilized in our analysis,
the ratio of $\approx 250$ is again recovered.

The accuracy of the nearest-neighbors strategy
(or at least the way we implemented the idea) is
lower than that of the binning approach;
see Fig.\ \ref{fig-bin}a.
However, the results in Fig.\ \ref{fig-nobin}
are encouraging, and demonstrate
that it is possible to estimate $\pobs$
using a non-binning procedure.

\subsection{Note on efficiency}\
It is worthwhile to briefly discuss the efficiency of the re-weighting
approach using Eq.\ \eqref{eq-bb-df} compared
with that of using Eq.\ \eqref{eq-can-df}.
There are, in general, two extreme cases to consider:
(i) The barrier between the states of interest is very low.
This implies that barrier crossing events occur frequently, and thus
counting via Eq.\ \eqref{eq-can-df} will likely be the most
efficient approach.
(ii) The barrier between the states is too high to cross during
ordinary simulation.
In this case, counting via Eq.\ \eqref{eq-can-df} is not possible.
Importantly however, the re-weighting approach of Eq.\ \eqref{eq-bb-df}
can still be utilized since all that is required
are independent simulations in each state.

The fairly lengthy simulations required for the BBRW method in
the case of dileucine reflects the choice of state definitions:
\emph{within each state}, i.e., alpha or beta,
the rotameric (side-chain) timescales are
substantially longer than those of the backbone torsions.

\subsection{Dimensionality}\
With dileucine we were able to obtain accurate results for $\pobs$,
and thus $\df$, using only 4 degrees of freedom---out of a possible 144.
The underlying assumption is that the the degrees
of freedom not included in the analysis will provide the same
contribution to each relative weight (see Supplementary Material for details).
Thus, at a minimum, the coordinates used to define the states
of interest must be included in the re-weighting analysis
(here, the backbone dihedrals $\Phi$ and $\Psi$).

A particularly promising strategy for the future is to use
principal components analysis, or related methods,
 to suggest an optimal reduced set of coordinates
\cite{karplus-deca,kavraki,clementi-dim}.
These collective coordinates can then be used for binning or
to determine configuration-space distances for the nearest-neighbor approach.

\section{Conclusions}
The black-box re-weighting (BBRW) strategy
computes ensemble averages for any target distribution based on 
estimating the observed configuration-space density actually sampled
in a simulation---without employing typical assumptions about sampling quality.
In principle, the approach can be used to re-weight any ensemble, regardless of
the process used to generate the configurations.
Our data show that using BBRW can dramatically reduce
both systematic and statistical error for some systems.

As a proof of principle, we used the approach to
estimate free energy differences ($\df$) for non-overlapping states
of one-dimensional and molecular systems, based on completely independent
ensembles generated in each state.
Further, when ordinary one-dimensional trajectories exhibiting
transitions between states were re-analyzed with BBRW, statistical
error was substantially reduced.

Practical application of BBRW hinges on estimation of the
observed configuration-space density, $\pobs$, but the
theoretical basis does not.
We have obtained reasonable results using two different methods
for estimating $\pobs$, emphasizing that 
\emph{the central idea behind our re-weighting approach 
is independent of the specific strategy used to estimate $\pobs$}.
Strategies beyond those examined in this report are available
\cite{silverman-book,scott-book} and will be explored in future work.

There are two primary limitations of the overall approach.
First, improved density estimation schemes will be needed for
higher dimensional systems.
Second, BBRW is an \emph{analysis} method that re-weights
configurations in \emph{existing} ensembles;
thus, it cannot predict populations for regions of
configuration space not represented in the original data.

Our motivation for the black-box approach stems from the
still-outstanding challenge of producing statistical ensembles for
full-sized proteins.
The black-box approach seems promising in this context because:
(i) Non-Boltzmann-distributed sets of diverse structures can be
generated using a variety of methods, including NMR structure
prediction software \cite{spronk}, and by the addition of
atomic detail to coarse models \cite{micheletti,rapper}.
(ii) Local canonical sampling based on \emph{ad hoc} starting
structures is readily possible with existing software.
(iii) Our own box-counting studies of proteins (unpublished)
suggest that folded proteins act as systems with dramatically
reduced dimensionality; see also \cite{kavraki,clementi-dim}.
We also hope that BBRW will prove useful in re-weighting implicitly
solvated biomolecules into explicit solvent ensembles.
Finally, the idea could prove of value in the analysis of data
from insufficiently converged canonical simulation of any type.

\subsection*{Acknowledgments}
We thank Edward Lyman, Robert Swendsen, and David Zuckerman
for very helpful discussions.
Funding was provided by the 
Idaho EPSCoR, by 
NIH grants GM076569, CA078039,
and fellowship GM073517,
as well as by NSF grant MCB-0643456.

\end{document}